\documentclass[twocolumn,floatfix,preprintnumbers, superscriptaddress, reprint, prl]{revtex4-1}
\usepackage[english]{babel}
\usepackage[utf8]{inputenc}
\usepackage[colorinlistoftodos, color=green!40, prependcaption]{todonotes}

\usepackage{amsmath}
\usepackage{stmaryrd}
\usepackage{txfonts}
\usepackage{amssymb}
\usepackage{mathrsfs}
\usepackage{dcolumn}
\usepackage{bm}
\usepackage{braket}
\usepackage{epsfig}
\usepackage{color}
\usepackage{ulem}
\usepackage{bibunits}

\usepackage{physics}
\usepackage{xcolor}
\usepackage{graphicx}
\usepackage[left=23mm,right=13mm,top=35mm,columnsep=15pt]{geometry} 
\usepackage{adjustbox}
\usepackage{placeins}
\usepackage[T1]{fontenc}
\usepackage{csquotes}

\usepackage[colorlinks=true, linkcolor=blue, citecolor=blue, urlcolor=blue]{hyperref}

\begin{document}
\title{Evidence for the Alternating Next-Nearest Neighbor model in the dynamic behavior of a frustrated antiferromagnet}

\author{Adra Carr}
\thanks{These authors contributed equally}
\affiliation{Center for Integrated Nanotechnologies, Los Alamos National Laboratory, Los Alamos, New Mexico 87545, USA}
\affiliation{National High Magnetic Field Laboratory, Los Alamos National Laboratory, Los Alamos, New Mexico 87545, USA}
\author{John Bowlan}
\thanks{These authors contributed equally}
\affiliation{C-PCS, Los Alamos National Laboratory, Los Alamos, NM 87545}
\author{Claudio Mazzoli}
\author{Andi Barbour}
\author{Wen Hu}
\author{Stuart Wilkins}
\affiliation{NSLS-II Brookhaven National Laboratory, Upton, New York, USA}
\author{Colby Walker}
\affiliation{Center for Integrated Nanotechnologies, Los Alamos National Laboratory, Los Alamos, New Mexico 87545, USA}
\affiliation{Department of Physics and Astronomy, Brigham Young University, Provo, Utah, 84602}

\author{Xiaxin Ding}
\altaffiliation[Now at: ]{Idaho National Laboratory, Idaho Falls, Idaho}
\affiliation{National High Magnetic Field Laboratory, Los Alamos National
 Laboratory, Los Alamos, New Mexico 87545, USA}
\author{Jong Hyuk Kim}
\author{Nara Lee}
\author{Young Jai Choi}
\affiliation{Dept of Physics, Yonsei University, Seoul 03722, Korea}

\author{Shi-Zeng Lin}
\affiliation{Theory division, Los Alamos National Laboratory, Los Alamos, New Mexico 87545, USA}
\author{Richard L. Sandberg}
\affiliation{Center for Integrated Nanotechnologies, Los Alamos National Laboratory, Los Alamos, New Mexico 87545, USA}
\affiliation{Department of Physics and Astronomy, Brigham Young University, Provo, Utah, 84602}
\author{Vivien S. Zapf}
\affiliation{National High Magnetic Field Laboratory, Los Alamos National Laboratory, Los Alamos, New Mexico 87545, USA}

\begin{abstract}

X-ray photon correlation spectroscopy (XPCS) enables us to study dynamics of antiferromagnets. Using coherent soft X-ray diffraction, we resonantly probe Mn and Co Bragg peaks in the frustrated magnetic chain compound Lu$_2$CoMnO$_6$ significantly below the N\'{e}el temperature. Bragg peaks of incommensurate order slide towards commensurate `up up down down' order with decreasing temperature. Antiferromagnetic inhomogeneties produce speckle within the Bragg peaks, whose dynamics are probed by XPCS and compared to the classic Axial Next-Nearest Neighbor Interaction model of frustration. The data supports a novel model prediction: with decreasing temperature the dynamics become faster.
\end{abstract}

\date{\today}
\maketitle

Ferromagnetic (FM) domains have been studied for more than a century, revealing their key role in controlling the magnetic switching behavior and many other properties relevant to ferromagnetic applications.  By contrast, antiferromagnetic (AFM) domains and other inhomogeneities have historically been challenging to image due to their lack of net magnetization. In the past few decades however, promising techniques have been developed based on X-rays \cite{Schwickert1998,Kortright1999,Nolting2000,Stohr2006,Zhao2006,Kim18_Nature} including X-ray photon correlation spectroscopy (XPCS). \cite{Shpyrko07,Konings11,Chen13,Chen2016} AFM Bragg peaks are observed at the resonant edge of magnetic elements, allowing us to resolve speckle in the Bragg peak from coherent scattering off of AFM domains and inhomogeneities. The autocorrelation function of this speckle can be analyzed to extract statistical information. XPCS has led to e.g. observations of the dynamics of spin density waves in Cr \cite{Shpyrko07} and of spin-helix phases in Dy \cite{Chen13, Konings11}. These techniques are timely to address the emergence of innovative ideas for the application of AFMs, such as exchange bias \cite{Nogues99}, spintronics \cite{Bader2010,barker2016static,Gross2017} and multiferroics \cite{Ramesh07,Chu08,Marti2014,Legrand2019} as well as fundamental questions about the role of dynamic AFM in frustrated magnets. Thus far, XPCS studies of antiferromagnets have found that when the samples are cooled slightly below the N\'{e}el temperature $T < T_N$, the AFM dynamics become slow or freeze, as would be intuitively expected from thermal activation \cite{Shpyrko07, Chen13, Konings11}.

Here we focus on an AFM material with dynamics over an extended range of $T$, due to strong geometrical frustration. The inability of frustrated magnets to simultaneously satisfy all pairwise interactions leads to many nearly-degenerate states near the ground state, and thus enables inhomogeneities in space and time. Frustration in AFM was theoretically proposed shortly after the proposal of AFM itself \cite{Neel48,Wannier50}. However, to date, the basic dynamic predictions have been difficult to test directly.  Thus, the recently-developed technique of XPCS tuned to the resonance of magnetic ions and the AFM Bragg peak is important to enable us to tackle the problem of AFM domain dynamics.

We study a long-standing problem in frustrated magnetism - the Axial Next-Nearest Neighbor Ising (ANNNI)  model \cite{Bak82, Selke88}. In this model, nearest-neighbor FM interactions compete with next-nearest-neighbor AFM interactions along chains of Ising spins. This apparently simple model predicts complex behavior. As an ANNNI system is cooled below $T_N$, it passes through a theoretically infinite number of first order phase boundaries separating AFM phases with different commensurate (CM) or incommensurate (ICM) wavevectors, referred to as the `Devil's staircase' \cite{bak_theory_1980}. This large density of phases and phase transitions below $T_N$ has interesting dynamic consequences, as we shall show both experimentally and theoretically below. At the lowest $T$, the ANNNI model predicts a CM wave vector such as `up up down down' magnetic moment orientation along the chains. There have been previous XPCS investigations of e.g. frustrated magnets Dy and Cr that also show ICM order due to frustration between nearest and next-nearest neighbors, however they do not have Ising spins and thus they are not examples of the ANNNI model and dynamics were only observed in the immediate vicinity of $T_N$ \cite{Shpyrko07,Chen13}.

\begin{figure}
\centering
\includegraphics[width=3 in]{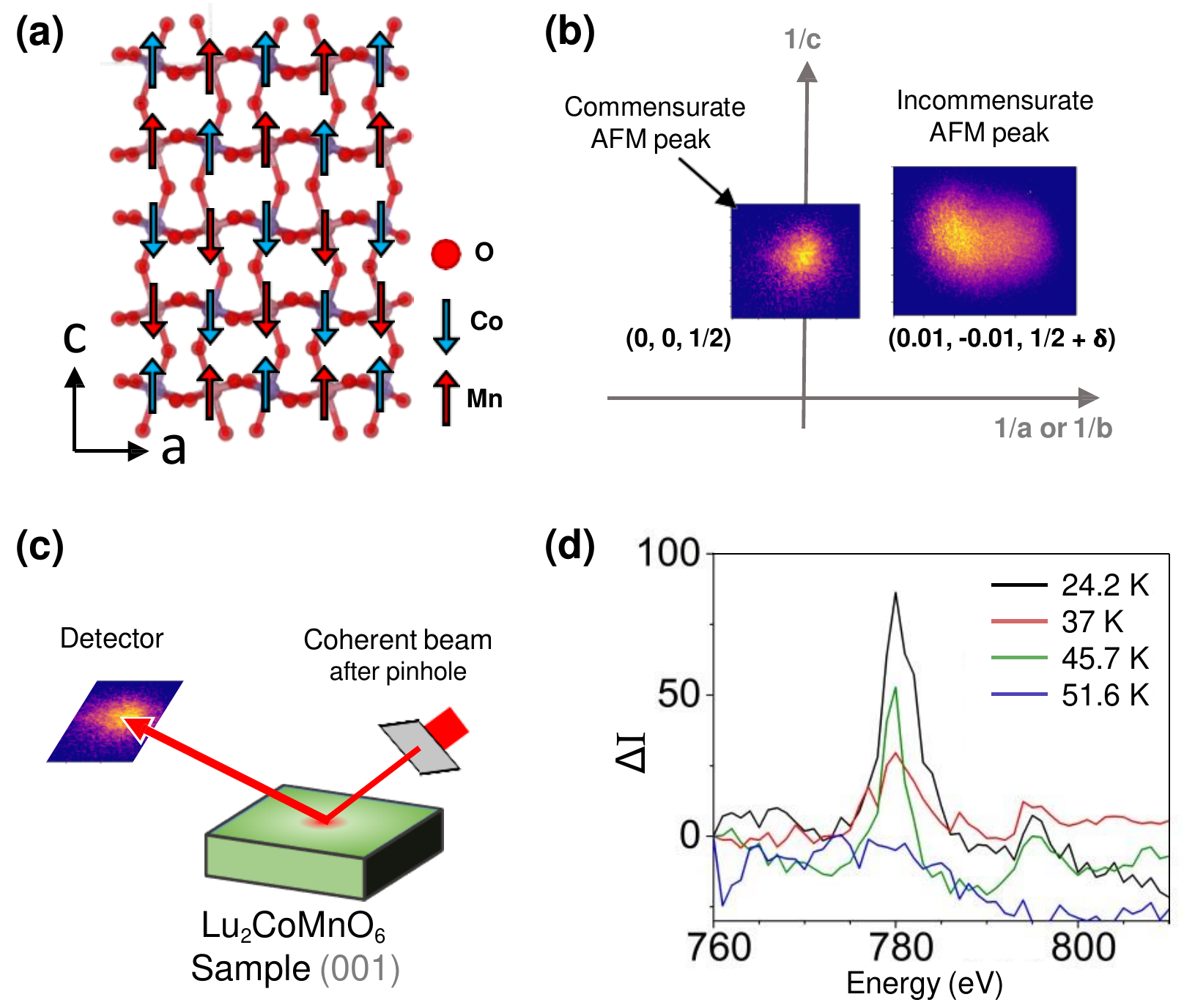}
\caption{(a) Structure of Lu$_2$CoMnO$_6$ with spins, shown as arrows, with Lu atoms omitted. (b) Detected CM and ICM Bragg peaks, shown with their respective reciprocal space wavevector. (c) Experimental setup of resonant XPCS experiment, where the $c$-axis is normal to the illuminated face of the sample. (d) Energy spectrum for the ICM Bragg peak at the Co $L_{3}$ edge. Peak intensity was integrated over the detector area, with $\Delta I$ expressed as the relative change of the intensity at 780 eV vs. incident X-ray energy.}

\label{experiment}
\end{figure}

We investigate the double perovskite compound Lu$_2$CoMnO$_6$ \cite{YanesVilar11,Lee14,Chikara16,Choi17}. Its physical properties, muon spin resonance ($\mu$sR) and neutron diffraction experiments have been previously interpreted in the context of the ANNNI model \cite{Zapf16}. Co$^{2+}$ and Mn$^{4+}$ with $S = 3/2$ spins occupy oxygen cages, and the two magnetic ions alternate along the a, b, and c axes \cite{YanesVilar11} with lattice spacing of $a$ = 5.1638(1) \AA, $b$ = 5.5467(1) \AA, $c$ = 7.4153(1) $\AA$. 
A powder neutron diffraction study found that the magnetic state at 4 K consists of  Co~$\uparrow$~Mn~$\uparrow$~Co~$\downarrow$~Mn~$\downarrow$ configurations with both the spins and wave vector along the $c$ axis, and a slight incommensuration with $\Vec{k} \approx [0.0223(8),0.0098(7),0.5]$ (see Fig. \ref{experiment}(a)) \cite{YanesVilar11}. Since the compound is multiferroic \cite{YanesVilar11,Lee14,Chikara16,Choi17} previous studies used the magnetization and electric polarization to determine that this system has slow dynamics on hour time scales, as well as frequency dependence of these quantities in the kHz frequency range. These dynamics were observed between $T_N = 48$ K and $T_{\rm Hyst} = 30-35$ K. Dynamics between ps and $\mu$s timescales were also uncovered by measuring $\mu$SR and neutron diffraction \cite{Zapf16}. Below $T_{\rm Hyst}$, hysteresis appeared in the magnetization and electric polarization suggesting that AFM and field-induced FM domains become pinned below this temperature by spin-lattice interactions \cite{YanesVilar11,Choi17}. 

We performed XPCS measurements on mm-sized single crystals of Lu$_2$CoMnO$_6$ \cite{Chikara16} at the Coherent Soft X-ray Scattering (CSX, 23-ID-1) beamline at NSLS-II \cite{Chen2016} during three different beam times. The crystals were polished on their (001) faces down to 0.3 $\mu$m, and mounted with the (001) face upward on a copper sample holder with silver paint. Coherent X-rays passed through a 10 $\mu$m pinhole, then resonantly Bragg scattered off the Co or Mn ions in the geometry shown in Fig. \ref{experiment}. Each speckle pattern was recorded at a fixed $T$ every 3.25 s for up to three hrs, after allowing for thermalization (approximately 1 hour). The dynamics of the speckle pattern, and thus the domain patterns which they encode, were analyzed by computing the autocorrelation function $g^{(2)}(\Vec{q}, \tau)$ of the speckle intensity $I(\Vec{q},t)$: $g^{(2)}(\Vec{q}, \tau) = {\langle I(\Vec{q}, t)I(\Vec{q}, t+\tau)\rangle}/{\langle I(\Vec{q}, t)\rangle^2}$. Here, the intensities $I(\Vec{q}, t)$ and $I(\Vec{q}, t+\tau)$ are extracted for a particular momentum vector $\Vec{q}$ at times $t$ and $t+\tau$, with $\tau$ being a delay time and angle brackets denoting time averaging.

We investigate the two satellite ICM Bragg peaks $\Vec{k} = [\pm\delta, \mp\delta, 1/2\mp\epsilon]$ and a CM Bragg peak $\Vec{k} = [0, 0, 1/2]$, where the CM peak corresponds to the `up up down down' ordering. $\delta$ and $\epsilon$ are in the range of 0.01 or lower and decrease with $T$ \cite{SI}. Data was taken at Co and Mn edges for $\sigma$ and $\pi$ X-ray polarization, and a complete XPCS data set was taken at the Co edge with $\pi$ polarization since it had the largest magnitude. The ICM wave vectors are slightly different than those found in polycrystalline neutron diffraction data \cite{YanesVilar11}, however they are more accurate since they are extracted from single crystals. They include incommensuration along the c-axis, which is a prediction of the ANNNI model.

We find that above $T_{\rm Hyst}$ (the region where physical properties have no hysteresis in $T$ or magnetic field)  the nominal CM $[0, 0, 1/2]$ Bragg peak has no resolvable energy dependence near the Co or Mn $L_{3}$ edges, showing that is not at a magnetic resonance. We conclude that it is dominated by the first harmonic of the X-ray beam diffracting off the [0,0,1] lattice peak. However, below $T_{\rm Hyst}$ it acquires a strong resonance at the energy of the Co and Mn $L_{3}$ edges thus this Bragg peak becomes dominated by the [0,0,1/2] magnetic peak. Meanwhile, the ICM peaks can be observed only at the Co and Mn $L_{3}$ edges at all $T < T_N$, both above and below $T_{\rm Hyst}$, proving their magnetic character. There is a small $T$-dependence of $\delta$ and $\epsilon$ (shown in the S.I. \cite{SI}) such that the ICM peaks approaches the CM $\Vec{k} = [0, 0, 1/2]$ position as $T$ is lowered from $T_N$ to $T_{\rm Hyst}$, as predicted for the ANNNI model. The fact that the CM peak abruptly acquires a magnetic component below $T_{\rm Hyst}$, may indicate that part of the sample evolves all the way to the CM order as $T$ is lowered, while another part of the sample remains trapped in a state described by an ICM wavevector. The wave vectors of the CM and ICM peaks were reproducible at different locations on the sample, though the ICM peaks were not always visible. 

\begin{figure}
\centering
\includegraphics[width=3 in]{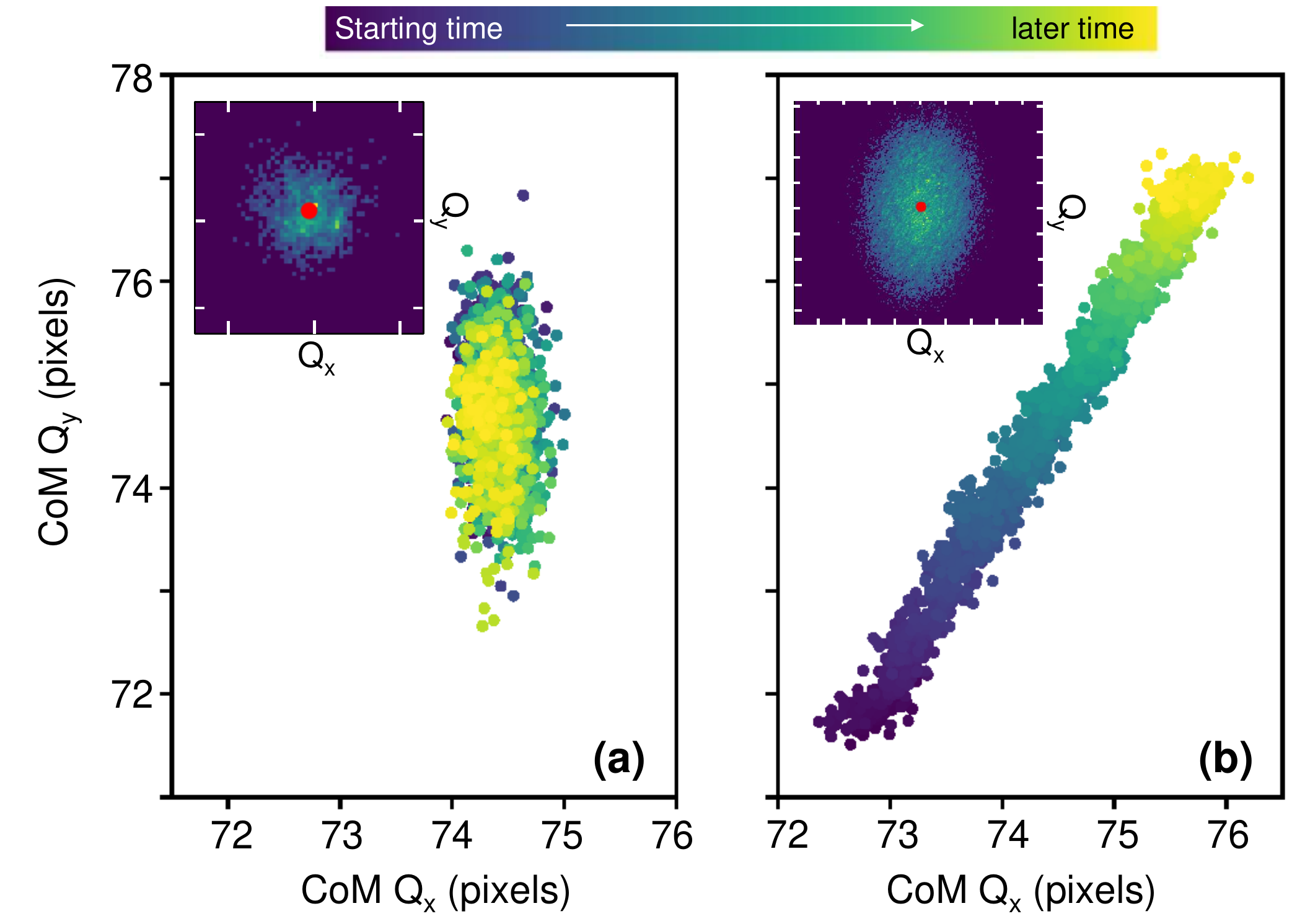}
\caption{Center of Mass (CoM) of CM (a) and ICM (b) peaks at 35 K shown in detector pixels (1 pixel = $7 \times 10^{-5} \AA^{-1}$). The color denotes relative position in time over 1.8 hrs. Insets show a full Bragg peak with red dot indicating CoM extracted from an Otsu thresholded detector image, and white hashmarks denoting 20 pixel intervals.}
\label{CoM_movement}
\end{figure}

We also noted a significant drift in the magnetic ICM Bragg peak position over time for $T>T_{Hyst}$. This drift is shown in Fig \ref{CoM_movement} over a roughly 2 hour period, with a lack of drift in CM peak shown for comparison. Additional information is given in the S.I. \cite{SI}. Similar drifts have been observed in the ANNNI spin chain compound Ca$_{3}$Co$_{2}$O$_{6}$ \cite{Moyoshi_2011, Agrestini_2011} and were attributed to the inability of the system to reach its stable ICM wave vector after a change in $T$ due to very slow dynamics.

\begin{figure}
\centering
\includegraphics[width=3 in]{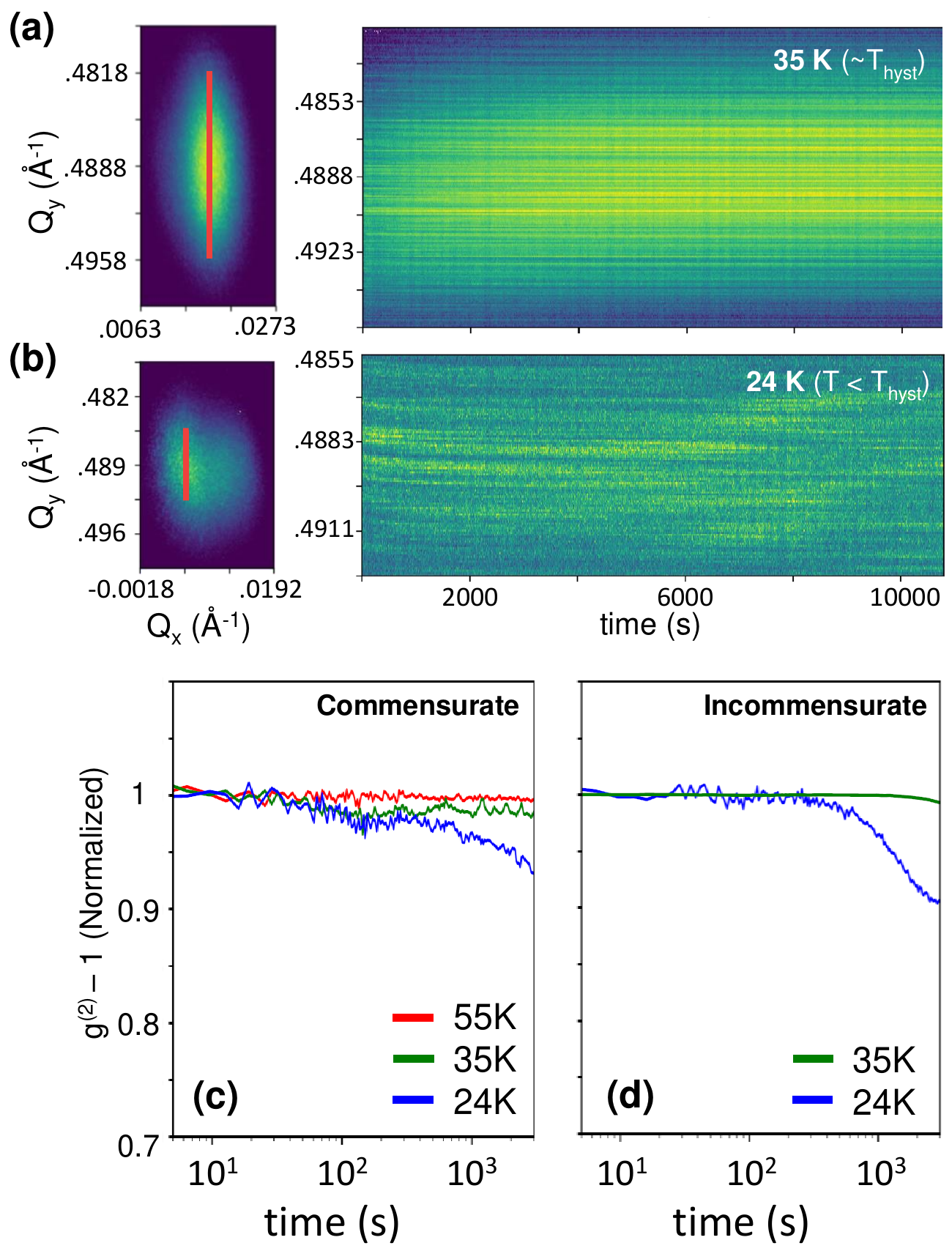}
\caption{a) (Right) Waterfall plot showing Cross section of ICM Bragg peak at 35 K shown vs time, where cross section is integrated over the red lineout defined on the Left. b) Similar waterfall plot for 24 K. c) Normalized autocorrelation function $(g^{(2)}-1)$ vs. time for the CM peak at [0, 0, 1/2] and d) ICM peak at [0.0087, -0.0042, 0.489] and [0.0168, -0.006, 0.4888] at 35 K. All data is shown for the Co $L_{3}$ edge with $\pi$ polarization.}
\label{speckle}
\end{figure}

Figure \ref{speckle} (a) and (b) shows the ICM peak at the Co edge at 778 eV, measured in a $\pi$ configuration at $T= 35$ K and 24 K, just above and below $T_{\rm Hyst} =$ 30 K. On the right are waterfall plots, showing an average over a vertical stripe through the center of the Bragg peak $1.4 \times 10^{-4} \AA^{-1}$ wide (2 pixels) vs time. The ICM speckle pattern is relatively unchanged at 35 K over the 3 hours, whereas at 25 K the speckle shifts and decorrelates on a shorter timescale. This behavior was verified in several different beam times. Thus, at 35 K the ICM peak drifts but does not decorrelate, and at 25 K the ICM peak decorrelates but does not drift.

In Fig. \ref{speckle} (c) and (d) the normalized intermediate scattering function $g^{(2)} - 1$ (the autocorrelation function between the signal at different times) is shown at $T =$ 25, 35 and 55 K. The autocorrelation is normalized to the average of $g^{(2)} - 1$ within the first 30 s of integration, serving as a representative baseline. Exact regions of interest (ROIs) used in the calculation can be found in the S.I. \cite{SI} Since 55 K is above $T_N$, there is no Bragg peak for the ICM case. Autocorrelation functions for other sub-regions of the CM Bragg peak were unable to discern notably separate dynamics between the central and outer portions of the Bragg peak (see S.I. \cite{SI}).

Thus, by means of XPCS we measure that the speckle from domains in the purely magnetic ICM peaks at 25 K (below $T_{\rm Hyst}$) decorrelate significantly more rapidly than at 35 K. While this observation is counterintuitive in general, since dynamics usually freeze at low $T$, it is in fact a prediction of the ANNNI model due to the presence of a Devil's staircase of 1$^{st}$ order phase transitions. These dense 1$^{st}$ order phase boundaries can lead to an effective pinning of ICM wavevectors and consequently slower magnetic dynamics. Below $T_{\rm Hyst}$ on the other hand, large stable magnetic domains form and the dynamic speckle behavior is produced by faster local fluctuations within each domain. In the following, we calculate the dynamics in an ANNNI Monte Carlo model. 


Monte Carlo simulations of the dynamic behavior of the ANNNI model were performed to compare with the experimental data. We note that in Lu$_2$CoMnO$_6$ there are two magnetic ions instead of one which is a deviation from the classic single-ion ANNNI model. We perform simulations for the two-ion ANNNI model (here) and compare to the single-ion ANNNI model (S.I.) to confirm that the dynamics are similar.

\begin{figure}[t]
\centering
\includegraphics[width=3 in]{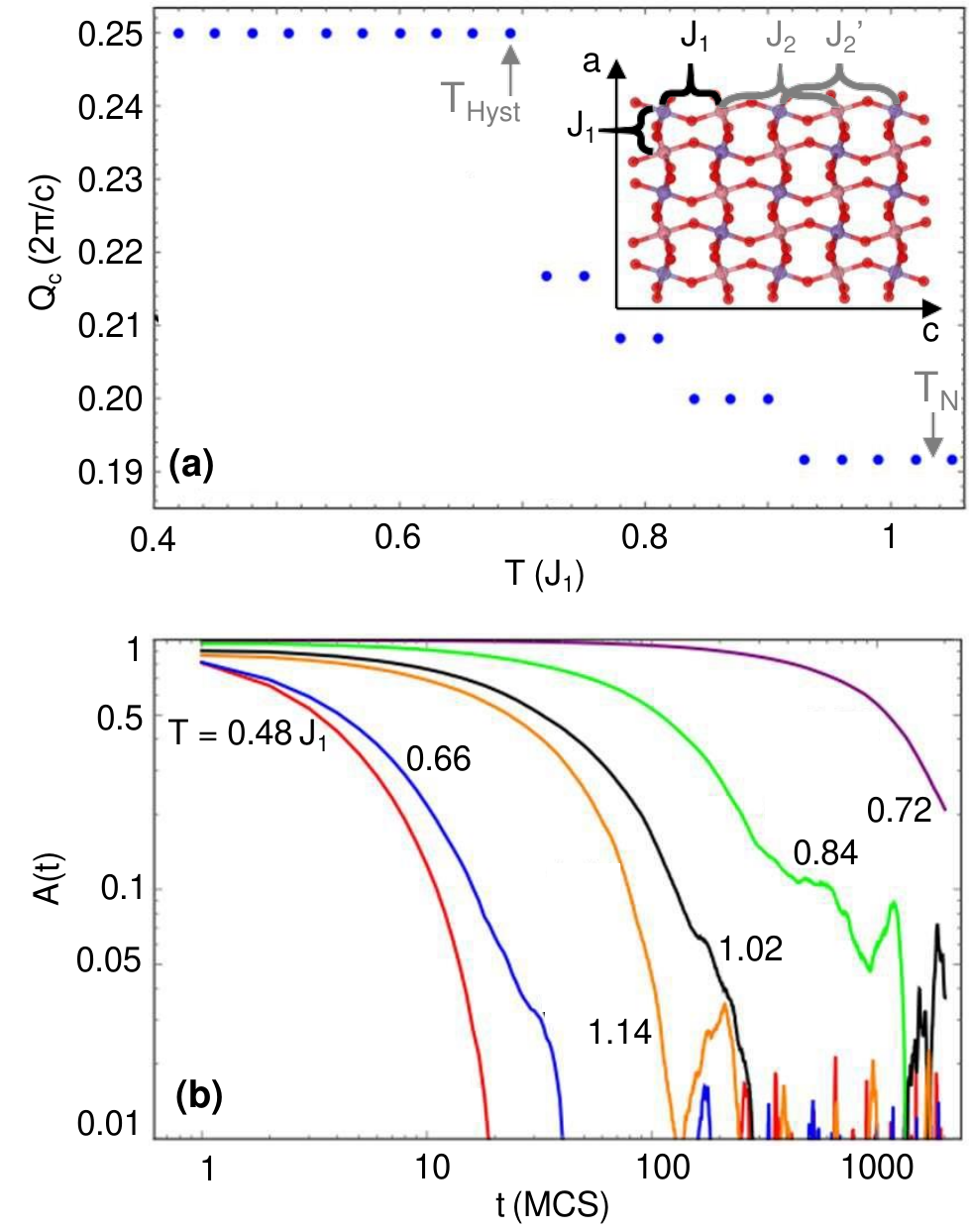}
\caption{(a) Monte Carlo simulations of the ICM AFM wavevector $Q$ vs $T$, obtained by slowly cooling to $T = 0$ and then heating. Inset shows the magnetic model with nearest and next-nearest neighbor magnetic interactions. (b) Calculated spin autocorrelation function $A(t)$ at fixed $Q=0.25\times 2\pi/c$ vs time for varying $T$. Time ($t$) is shown in units of Monte Carlo sweep (MCS).}                                                                                                                                                          
\label{fT1}
\end{figure}

The in-plane ordering wavevector is small, thus we assume a simple nearest-neighbor FM interaction in the $a-b$ plane. The spin Hamiltonian is expressed as:
\begin{align}\nonumber
    \mathcal{H}=-J_1 \sum _{NN} \sigma _i S_j^z-J_2 \sum _{NNN} \sigma _i \sigma _j-J_2' \sum _{NNN} \mathbf{S}_i\cdot \mathbf{S}_j-A \sum _i \left(S_i^z\right)^2,
\end{align}
where $\sigma_j=\pm 1$ is the Co$^{2+}$ Ising spin and $|\mathbf{S_j}|=1$ is the Mn$^{4+}$ Heisenberg spin with an easy axis anisotropy, $A>0$. $\mathbf{S_j}$ and $\sigma_j$ form two sublattices of the square lattice, as shown in Fig. \ref{fT1} (a).   $J_1>0$ is the nearest-neighbor (NN) FM interaction, and $J_2<0$ and $J_2'<0$ are the next-nearest neighbor (NNN) AFM interaction along the c axis. We choose $J_2=J_2'=-0.6 J_1$ and $A=J_1$ to match the experimental $T_N$ and $T_{\rm Hyst}$. We employ a 2-D model corresponding to the $a-c$ plane of Lu$_2$CoMnO$_6$ and perform Monte Carlo simulations of $\mathcal{H}$ with the standard Metropolis algorithm.    

The calculated ordering wavevector $Q$ vs. $T$ is shown in Fig. \ref{fT1}(a). Because of the finite size effect, $Q$ changes stepwise with $T$. There are peaks in the calculated specific heat when $Q$ changes, which correspond to the 1$^{st}$ order phase transition between different $Q$ states. The 1$^{st}$ order phase transition implies slow dynamics in the ICM phase. For increasing system size, $Q$ changes quasi-continuously through many weak 1$^{st}$ order transitions. To capture the dynamics, we compute the autocorrelation function of the spin structure factor $S(Q, t)\equiv \langle M_z(Q, t) M_z(-Q, t)\rangle$, represented as:
\[
A (t)={\int_0^\tau dt_1 \left(\mathcal{S}(t_1)-\bar{\mathcal{S}}\right) \left(\mathcal{S} \left(t+t_1\right)-\bar{\mathcal{S}}\right)}/{\int_0^\tau dt_1 \left(\mathcal{S}(t_1)-\bar{\mathcal{S}}\right)^2},
\]
where $M_z$ is the $z$ component of the magnetization at wavevector $Q$ and $\langle \cdots \rangle$ denotes thermal average. Here $t$ is the Monte Carlo time and $\tau$ is the total simulation time, and $\bar{\mathcal{S}}$ is the mean value of $S(Q, t)$. The results, displayed in Fig. \ref{fT1}(b), show that the dynamics for $T$ between $T_N$ and $T_{\rm Hyst}$ are extremely slow, slower than the region below $T_{\rm Hyst}$. The correlation time is particular long at the first order phase transition point.

The simulations of the dynamics are qualitatively consistent with the XPCS results. The theory and experiment both show `inverted' dynamics, where the speckle decorrelates faster below $T_{\rm Hyst}$ than above it. 

In conclusion, we observe dynamics of speckle in CM and ICM peaks over a very broad range of temperature down to a quarter of $T_N$, and these dynamics are inverted in temperature from ordinary magnets - we observe fast dynamics at low $T$ and slow dynamics at high $T$. We employ Monte Carlo simulations to show that these unusual dynamics are predicted by the ANNNI model. Above $T_{\rm Hyst}$ the ANNNI model predicts many (theoretically infinite) first order phase transitions occurring at closely-spaced temperatures whose domain boundaries pin the ICM wave vectors, creating slow dynamics. This manifests as slow decorrelation in the speckle measured by XPCS as well as drifting ICM wave vectors that do not stabilize within three hrs after changes in $T$. On the other hand, below $T_{\rm Hyst}$ the system has stable pinned domains and so the dynamics are dominated by fast fluctuations within each domain. 

\section{acknowledgements}
We are grateful to Ashish Tripathi, Benjamin Pound, Shalinee Chikara for helpful discussions and experimental aid. Scientific work was supported by the LDRD program at LANL. This research utilized beamline 23-ID-1 of the National Synchrotron Light Source II, a U.S. DOE Office of Science User Facility operated for the DOE Office of Science by Brookhaven National Laboratory under Contract No. DE-SC0012704; as well as the facilities of the Center for Integrated Nanotechnologies, an Office of Science User Facility operated for the U.S. DOE Office of Science by LANL; and finally the facilities of the National High Magnetic Field Lab, funded by the U.S. National Science Foundation through Cooperative Grant No. DMR-1157490, the State of Florida, and the U.S. DOE. Sample growth efforts at Yonsei University were supported by the National Research Foundation of Korea Grants NRF-2017R1A5A1014862 (SRC program: vdWMRC center, NRF-2018R1C1B6006859, and NRF-2019R1A2C2002601).

\bibliography{references}

\end{document}


\title{Supplemental Material: Evidence for the Alternating Next-Nearest Neighbor model in the dynamic behavior of a frustrated antiferromagnet}

\author{Adra Carr}
\thanks{These authors contributed equally}
\affiliation{Center for Integrated Nanotechnologies, Los Alamos National Laboratory, Los Alamos, New Mexico 87545, USA}
\affiliation{National High Magnetic Field Laboratory, Los Alamos National Laboratory, Los Alamos, New Mexico 87545, USA}
\author{John Bowlan}
\thanks{These authors contributed equally}
\affiliation{C-PCS, Los Alamos National Laboratory, Los Alamos, NM 87545}
\author{Claudio Mazzoli}
\author{Andi Barbour}
\author{Wen Hu}
\author{Stuart Wilkins}
\affiliation{NSLS-II Brookhaven National Laboratory, Upton, New York, USA}
\author{Colby Walker}
\affiliation{Center for Integrated Nanotechnologies, Los Alamos National Laboratory, Los Alamos, New Mexico 87545, USA}
\affiliation{Department of Physics and Astronomy, Brigham Young University, Provo, Utah, 84660}

\author{Xiaxin Ding}
\altaffiliation[Now at: ]{Idaho National Laboratory, Idaho Falls, Idaho}
\affiliation{National High Magnetic Field Laboratory, Los Alamos National
 Laboratory, Los Alamos, New Mexico 87545, USA}
\author{Jong Hyuk Kim}
\author{Nara Lee}
\author{Young Jai Choi}
\affiliation{Dept of Physics, Yonsei University, Seoul 03722, Korea}

\author{Shi-Zeng Lin}
\affiliation{Theory division, Los Alamos National Laboratory, Los Alamos, New Mexico 87545, USA}
\author{Richard L. Sandberg}
\affiliation{Center for Integrated Nanotechnologies, Los Alamos National Laboratory, Los Alamos, New Mexico 87545, USA}
\affiliation{Department of Physics and Astronomy, Brigham Young University, Provo, Utah, 84602}
\author{Vivien S. Zapf}
\affiliation{National High Magnetic Field Laboratory, Los Alamos National Laboratory, Los Alamos, New Mexico 87545, USA}

\maketitle

\beginsupplement
\section{Initial Sample Characterization}

\begin{figure}[t]
\centering
\includegraphics[width=\columnwidth]{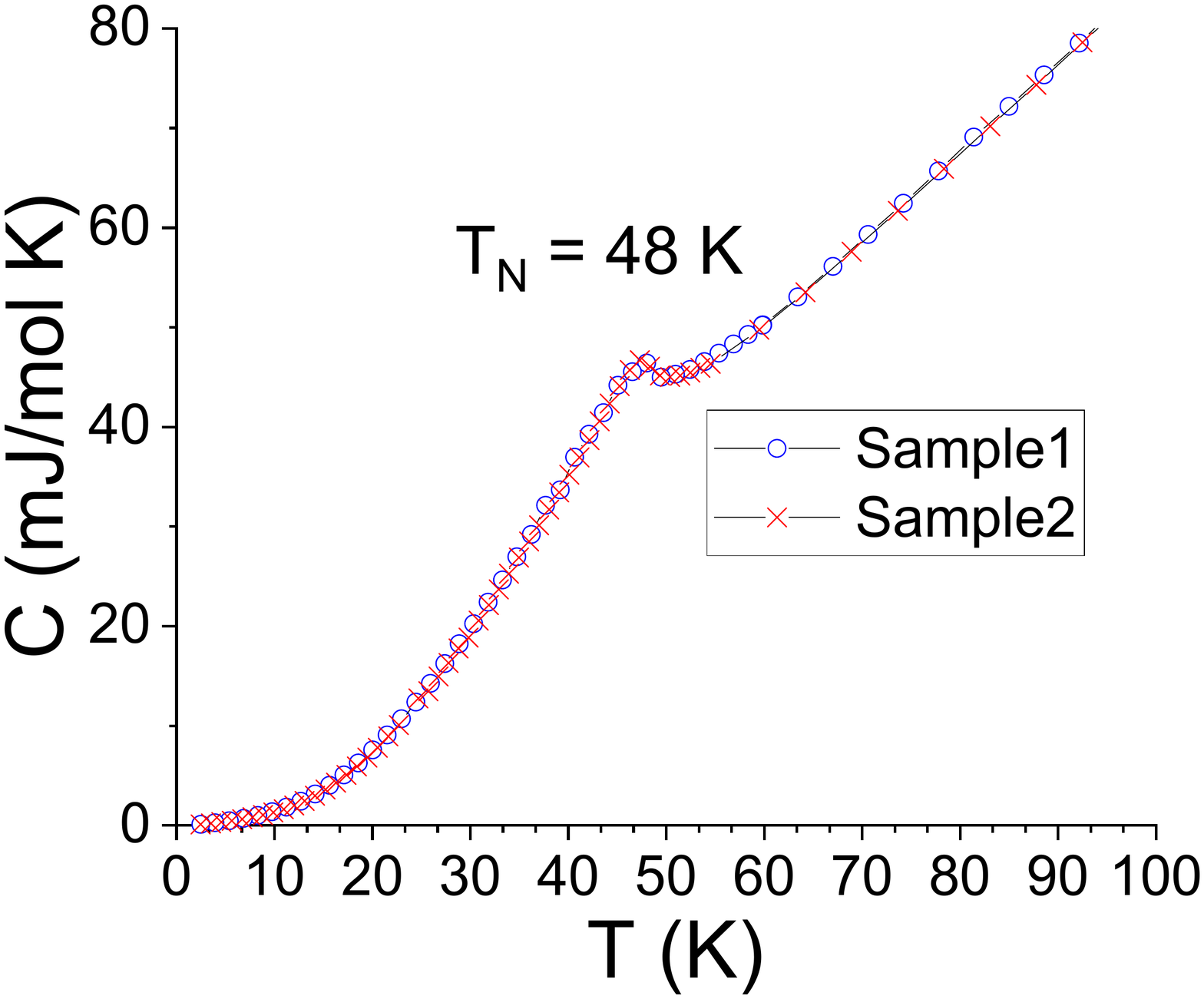}
\caption{Heat capacity is shown vs temperature measured in a Quantum Design Physical Properties Measurement System for samples 1 and 2. The two samples show consistent behavior with sharp peaks at $T_N = 48$ K.}
\label{Heatcapacity}
\end{figure}

Samples were grown by the flux method at Yonsei University \cite{Lee14}. Two millimeter-sized single crystals were measured as part of this study, from the same batch as those previously reported \cite{Lee14,Chikara16}. Samples were manually polished using 0.3 $\mu$m grit before measurements. Heat capacity was measured in a Physical Properties Measurement System (PPMS) between 4 and 100 K to verify the magnetic ordering below 48 K, as shown in Fig. \ref{Heatcapacity}.  One sample (S1) was used in all the analysis within the main text. A second sample (S2) was measured and showed differences in the specific ICM peak location, but the same dynamic behavior as a function of $T$.

\begin{figure}[t]
\centering
\includegraphics[width=\columnwidth]{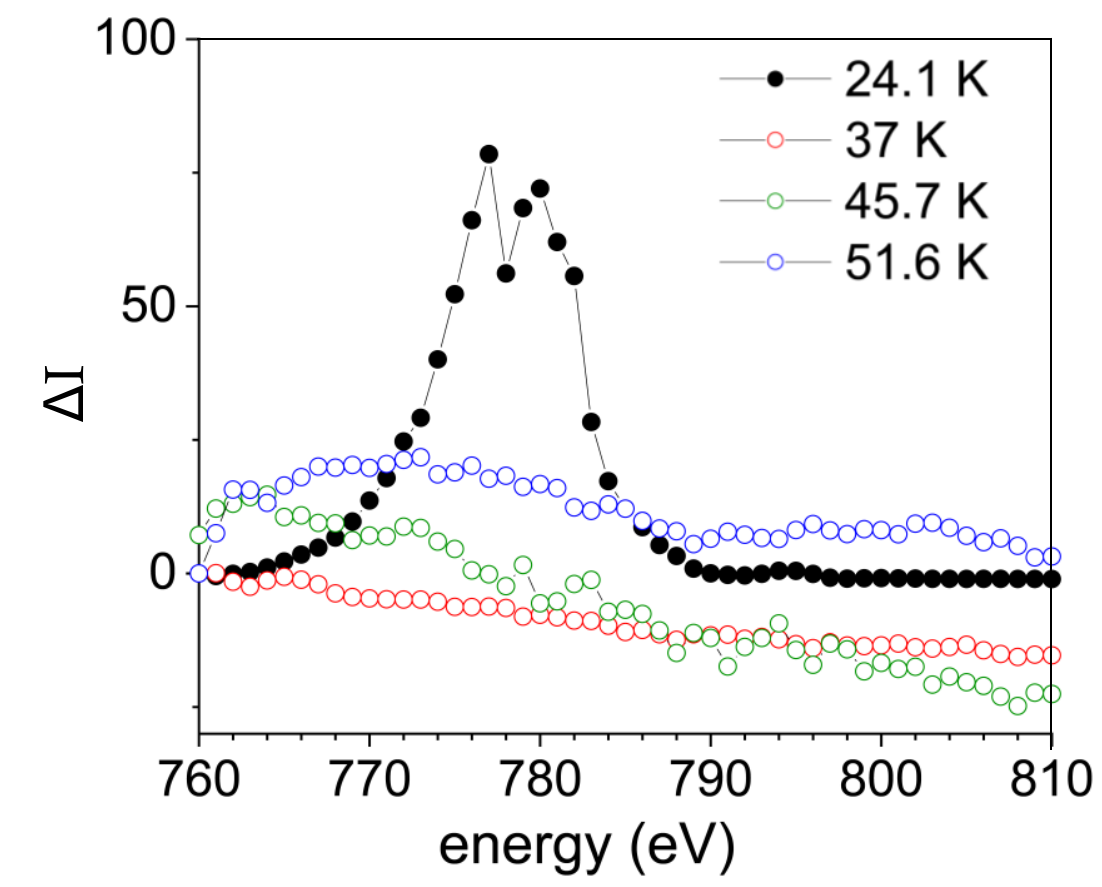}
\caption{Energy spectrum for the CM Bragg peak at the Co $L_{3}$ edge. Peak intensity was integrated over the detector area, with $\Delta I$ expressed as the relative change of the intensity at 780 eV vs incident X-ray energy. Scattering angles were adjusted to hold the scattering vector constant. Data is shown for Sample 1 with $\sigma$ polarization for the CM [0, 0, 0.5] Bragg peak.}
\label{CoSigmaEscan}
\end{figure}

Energy spectra for the CM and ICM Bragg peaks were collected at the NSLS-II beamline to verify the relative intensity increase due to the magnetic resonance at the Co $L_3$ and Mn $L_3$ edges. The ICM Co $L_3$ energy spectra appears in the main text, with the CM Co $L_3$ energy spectra shown in Fig \ref{CoSigmaEscan} for completeness. Intensity enhancement is primarily seen within the $T=$ 24 K regime, compared to near continuous enhancement below 50 K for the ICM peak in the main text. 

Energy spectra were also collected for the CM and ICM Bragg peaks at the Mn $L_3$ and $L_2$ edge, shown in Fig \ref{MnSigmaEscan}. Similar in behavior to the Co $L_3$ edge, intensity enhancement for the CM peak is only seen at the lowest temperatures, with enhancement seen in the $T < 55$ K regime for the ICM peak. All XPCS experiments within the main text were performed at the Co $L_3$ edge since it gave the strongest intensity enhancement for the ICM peak, making autocorrelation analysis of the speckle easier. 

\begin{figure}
\centering
\includegraphics[width=\columnwidth]{Mn_Sigma_energy_scan_sample 1_v2.pdf}
\caption{Energy spectrum for the (a) CM and (b) ICM Bragg peaks at Mn $L_{3}$ and $L_2$ edges. Similar to Figure \ref{CoSigmaEscan}, peak intensity was integrated over the detector area, with $\Delta I$ expressed as the relative change of the intensity at 645 eV vs incident X-ray energy. Data is shown for Sample 1 with $\sigma$ polarization, at indicated temperatures.}
\label{MnSigmaEscan}
\end{figure}

Additionally, movies have been included showing the speckle pattern of the ICM Bragg peak as a function of temperature upon cooling (\textit{Tdown S1.mov}) from 55 K and heating from 24 K (\textit{Tup S2.mov}). Total temperature range shown is between 55 and 25 K, with a heating/cooling rate of $ \sim$ 1 K/min and one frame corresponding to 6 seconds. Data shown is for Sample 1, at the Co $L_3$ edge. Detector x and y axes are shown in terms of pixel units, with one pixel width corresponding to $7 \times 10^{-5} \AA^{-1}$. Bad detector regions and a circular beam block, with the ability to mask the zero-order beam, have been masked as white.

\section{ROI for Commensurate and Incommensurate Bragg Peak $g^{(2)}$ Calculation}

\begin{figure}
\centering
\includegraphics[width=3.5in]{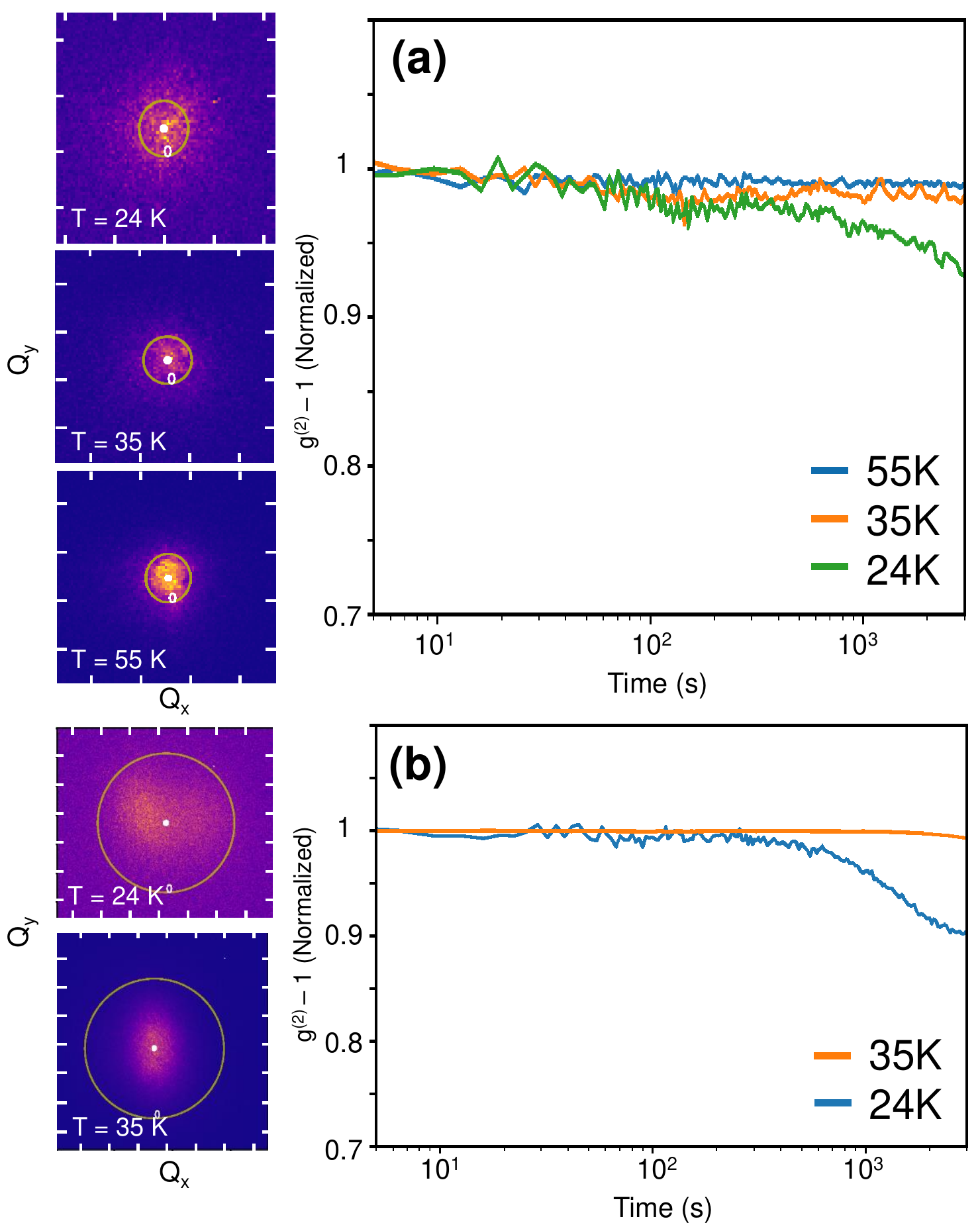}
\caption{(Left) Circular ROIs used in calculating the normalized $g^{(2)} - 1$ for the CM (a) and ICM (b) peaks and each temperature of interest. ROIs are outlined in yellow (with a region ``0" label), with white hashmarks denoting 20 pixel intervals and 50 pixel intervals for the CM and ICM peaks, respectively.}
\label{ROI_g2}
\end{figure}

Circular region of interests (ROIs) of differing $q$-vector size were used in the autocorrelation analysis seen in the main text. To have the ROI be well-centered on the Bragg peak, the center of the ROI was defined by first using Otsu's thresholding method to mask only relevant pixel values above background. \cite{Otsu79} Bad pixels on the detector image were additionally removed. The center of mass point was then calculated and defined as the center point for circular ROIs. The ROI radii were defined as $7 \times 10^{-4} \AA^{-1}$ (10 pixels) and $8.4 \times 10^{-3} \AA^{-1}$ (120 pixels) for the CM and ICM peaks, respectively. Shown in Fig \ref{ROI_g2}, ROI regions used in the $g^{(2)}$ calculation have been explicitly overlaid on detector images (yellow circles), with the normalized $g^{(2)} - 1$ plots reproduced from the main text for reference. 

\section{Autocorrelation of Commensurate Lattice Peak: Ring Regions of Interest}

\begin{figure*}[t]
\centering
\includegraphics[width=\textwidth]{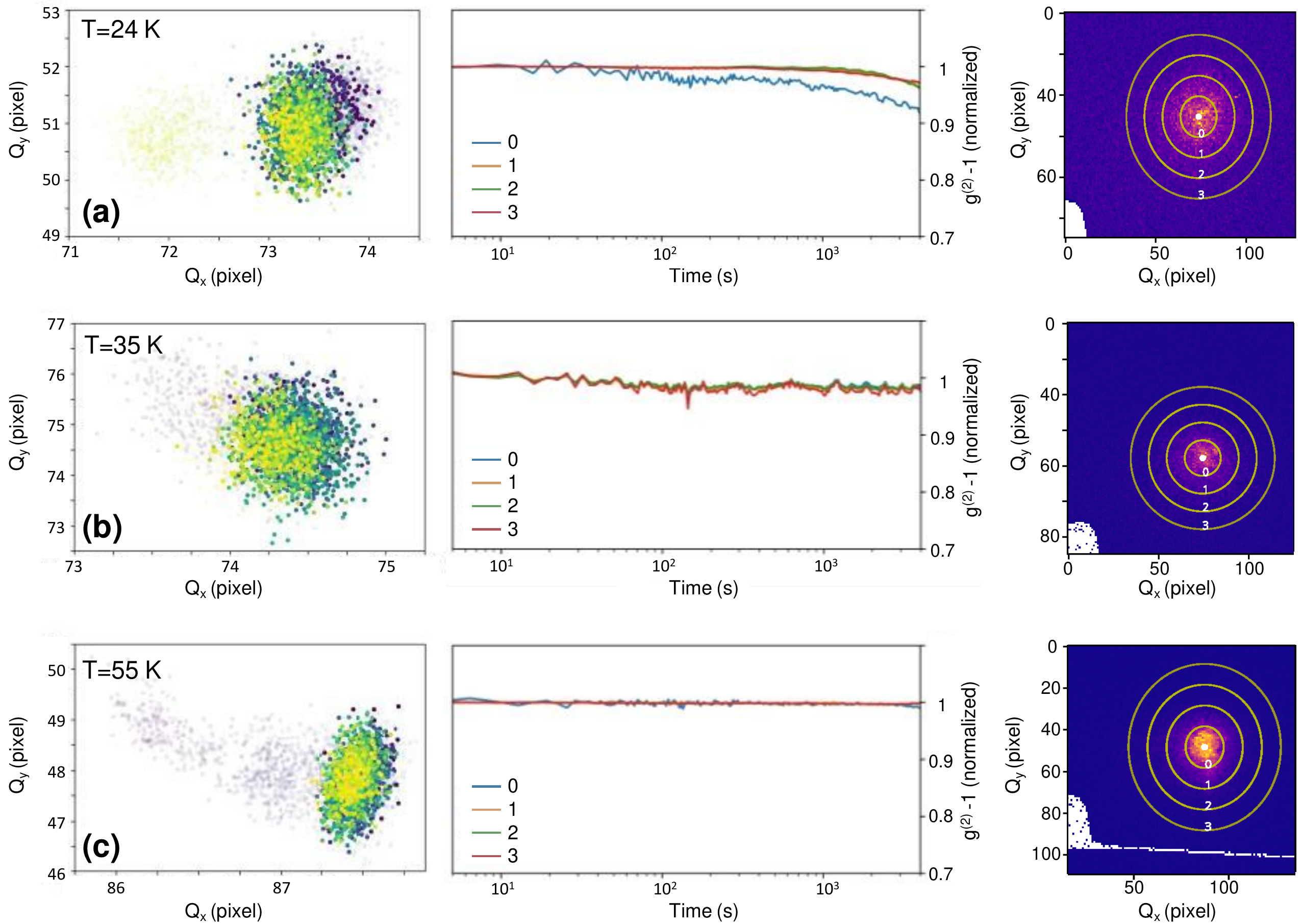}
\caption{Comparison of the autocorrelation time at temperatures of 24 K, 35 K, and 55 K, for 4 ring ROIs about the $[0, 0, 1/2]$ Commensurate lattice peak.  Data shown is for sample 1 at the Co $L_3$ edge, $\pi$ polarization. Masked detector pixels are shown as white.}
\label{ROI_ring}
\end{figure*}

In some magnetic systems, the scattering vectors for lattice and magnetic Bragg peaks lie on top of each other. However, due to the shorter correlation length of magnetic order, the magnetic Bragg peak will be broader and the lattice peak will be sharper. Thus, in a ``mixed" Bragg peak, the center of the peak can be dominated by scattering from the lattice and the periphery by magnetic scattering. In Lu$_2$CoMnO$_6$, all peaks studied occur at wavevectors corresponding to magnetic ordering. However, the CM [0,0,1/2] peak could have a contribution of the [0,0,1] lattice peak from the first harmonic of the X-ray beam. Thus, we do find a Bragg peak at [0,0,1/2] above $T_N$. Above $T_N$, this CM Bragg peak shows no resonance at the Co or Mn edges consistent with it coming from the [0,0,1] peak in the doubled frequency of the first harmonic of the X-ray beam (Fig. \ref{CoSigmaEscan}). For reasons not well understood, this peak also remains dominated by the lattice peak, e.g. not showing resonance in the energy scans, at 35 K below $T_{N}$ but above $T_{Hyst}$. The energy scan does show resonance and a dominance of the magnetic peak below $T_{Hyst}$.

Thus, the question arises whether the CM Bragg peak at 25 K and 35 K is a hybrid peak containing magnetic and lattice contributions, where the lattice dominates at the center of the peak and the magnetic contribution near the periphery. It is difficult to compare the intensities of the 25, 35 and 55 K peaks since the location on the sample varies due to thermal contraction of the experimental apparatus. We thus test if there is a difference in dynamics between the center of the Bragg peak and the periphery. We show the difference in decorrelation time between the inner and outer portions of the CM peak in Fig \ref{ROI_ring}. Ring ROIs were defined about the peak CoM and the autocorrelation for each region was extracted. A portion of the total data collected was used in the autocorrelation calculation to ensure the sample was in a ``well-thermalized" regime (shown at left for each temperature of interest). This was judged from the CoM extraction as the time period where there was minimal drifting in the peak center, shown with the same colorscale as Fig. 2 within the main text to denote relative position of the peak in time. Data not used in the calculation are shown as semi-transparent, with points at full opacity corresponding to a total time span of $3 \times 10^3$ s. The width of each ROI ring was defined as $7 \times 10^{-4} \AA^{-1}$ (10 pixels) and is explicitly shown as yellow rings with corresponding region labels of ``0" through ``3" (right). The normalized $g^{(2)} - 1$ for each temperature and each ROI is shown at center. 

At 55 and 35 K, there is no dependence of the decorrelation on the distance from the center of the peak. At 24 K, the innermost portion of the Bragg peak decorrelates fastest. Thus we do not observe any data consistent with a well-correlated center of the Bragg peak resulting from a lattice contribution. 

\section{Calculations for the single-ion ANNNI model}

In the main text, we show calculations for a two-ion ANNNI model. Here, we show similar calculations for a single-ion ANNNI model, and find similar dynamics.

The spin Hamiltonian reads  
\[
\mathcal{H}=-\sum _{\langle \text{ij}\rangle }  J_{{ij}}\sigma _i \sigma _j,
\]
where $\sigma_j=\pm 1$ is the Ising spin. We consider the 2D model with $J_{ij}=J_1>0$ for the nearest neighbor interaction and $J_{ij}=J_2<0$ for the next nearest neighbor interaction only in the $y$ direction. The spins order ferromagnetically in the $x$ direction and can be modulated in the $y$ direction. The ``up-up-down-down" state of the spins modulated in the $y$ direction is favored at $T=0$ when $-J_2/J_1>0.5$. Here we choose $J_2=-0.6 J_1$.   

We perform Monte Carlo simulations of the $2$D model with $\mathbf{Q}$ in the $y$ direction. The system size is $120\times 120$ lattice spins sites. The equilibrium phase diagram was studied in Refs. \cite{Selke1980,PhysRevB.81.094425,PhysRevB.90.144410,PhysRevB.95.174409}, with some controversy about the exact transition temperature. We take extra effort to equilibriate the system and use two protocols: 
\begin{enumerate}

\item Slowly cooling the system from $T=3 J_1$ to the target temperature with $2\times 10^7$ Monte Carlo sweeps (MCS) and $4\times 10^7$ MCS to thermalize the system and another $2\times 10^7$ MCS to perform the measurement. 

\item We prepare the system at $T = 0$ in the ``up-up-down-down" state and then instantly heat up the system to the target temperature. We use $4\times 10^7$ MCS to thermalize the system and another $2\times 10^7$ MCS to perform the measurement.

\end{enumerate}

\begin{figure}[t]
\centering
\includegraphics[width=\columnwidth]{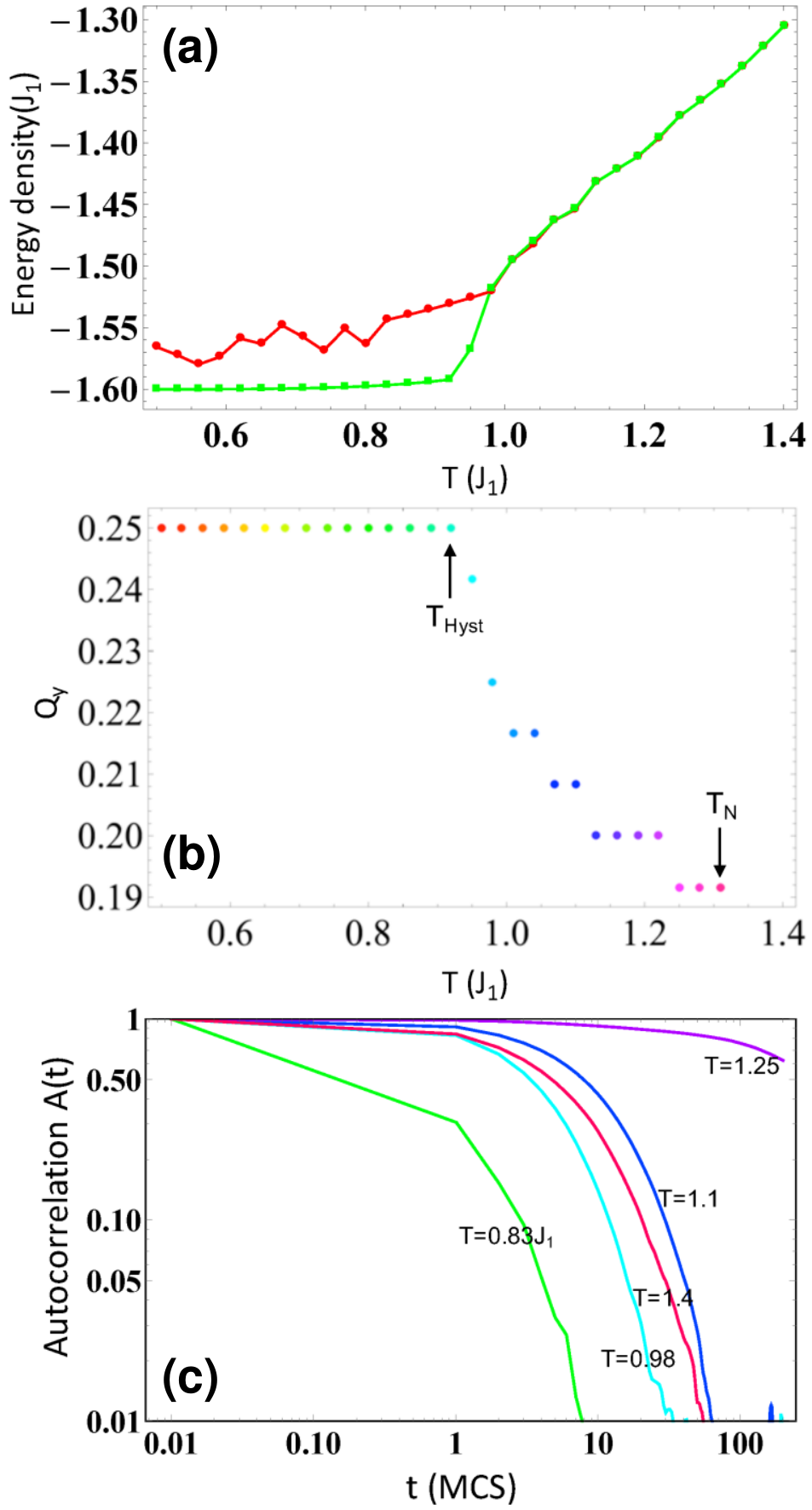}
\caption{Monte Carlo simulation results for the single-ion ANNNI model. (a) Energy vs. $T$ obtained by annealing (red) and heating (green). (b) $Q_y$ vs $T$ obtained by heating the system gradually. The color of the symbols match the color of lines for $A(t)$ (c) Calculated Autocorrelation $A(t)$ at $Q=0.25\times 2\pi/c$ vs time for different $T$, in units of Monte Carlo sweep (MCS).}
\label{fT1}
\end{figure}

As shown Fig.\ref{fT1}(a), when we slowly cool the system through the temperature range between $T_{N}$ and $T_{\rm Hyst}$, the system becomes trapped at an ICM AFM wave vector and never reaches the CM 'up up down down' state at low $T$ (red line). The green line is obtained by using protocol (2). The result of $Q$ vs. $T$ by using protocol (2) is shown in Fig. \ref{fT1}(b). Because of the finite size effect, $Q$ changes stepwise with $T$. There are peaks in the specific heat when $Q$ changes, which corresponds to the $1^{\rm st}$ order phase transition between different $Q$ states. The $1^{\rm st}$ order phase transition implies slow dynamics in the ICM phase. To capture the dynamics, we compute the autocorrelation function of the spin structure factor $S(Q, t)$ at $Q=(0, 0.25)$ defined in the main text.

The results are displayed in Fig. \ref{fT1}(c). As we can see, the dynamics {between $T_N$ and $T_{\rm Hyst}$} are extremely slow - slower than the region below $T_{\rm Hyst}$. The correlation time is particularly long at the $1^{\rm st}$ order phase transition point. In the infinite system size limit, the $1^{\rm st}$ order transition can occur at arbitrary small $Q$ interval, except for a small plateau of $Q$. The simulations are consistent with the two-ion model discussed in the main text. Our results suggest that the slower dynamics between $T_N$ and $T_{hyst}$ is rather universal in strongly frustrated systems.

\bibliography{references}